\documentstyle[12pt,epsf]{article}
\newcommand{\gtap}{\;{\raise.3ex\hbox{$>$\kern-.75em\lower1ex\hbox{$\sim$}}}\;}
\newcommand{\ltap}{\;{\raise.3ex\hbox{$<$\kern-.75em\lower1ex\hbox{$\sim$}}}\;}

\begin{document}

\begin{titlepage}

\hspace*{\fill}\parbox[t]{4.4cm}
{hep-ph/0202205 \\
Fermilab-Pub-01/389-T \\
ILL-(TH)-02-1 \\ \\
\today} \vskip2cm
\begin{center}
{\Large \bf Measuring the top-quark Yukawa coupling \\[7pt]
at hadron colliders via $t\bar th$, $h\to W^+W^-$} \\
\medskip
\bigskip\bigskip\bigskip
{\large  {\bf F.~Maltoni}$^1$,
         {\bf D.~Rainwater}$^2$,
     and {\bf S.~Willenbrock}$^1$} \\
\bigskip
$^{1}$ Department of Physics, University of Illinois at Urbana-Champaign \\
1110 West Green Street, Urbana, IL\ \ 61801 \\
\medskip
$^{2}$Theoretical Physics Department, Fermi National Accelerator Laboratory \\
Batavia, IL\ \ 60510 \\
\end{center}
\bigskip\bigskip\bigskip

\vspace{.5cm}

\begin{abstract}
We study the signal and background for the process $t\bar th$, $h\to W^+W^-$
at the LHC and a 100 TeV VLHC.  Signals are studied in two-, three-, and
four-lepton final states. We find a statistical uncertainty in the top-quark
Yukawa coupling at the LHC of $16\%, 8\%, 12\%$ for $m_h = 130, 160, 190$ GeV,
respectively.  The statistical uncertainty at the VLHC is likely to be
negligible in comparison with the systematic uncertainty.
\end{abstract}

\vfil

\end{titlepage}

\section{Introduction}
\label{sec:intro}

One of the primary quests of present and future colliders is the search for
the Higgs boson.  Precision electroweak data indicate that the mass of the
Higgs boson lies in the range 114 GeV $< m_h \ltap$ 200 GeV
\cite{:2001xw,Abbaneo:2001ix}. Such a Higgs boson may be discovered in Run II
of the Fermilab Tevatron \cite{Carena:2000yx}, and it cannot escape the CERN
Large Hadron Collider (LHC) \cite{ATLAS,CMS}.

Once a Higgs boson is discovered, it will be important to measure its couplings
to other particles to establish whether these couplings are those of a
standard-model Higgs boson, or those of a Higgs boson from an extended Higgs
sector, such as a two-Higgs-doublet model.  For example, the minimal
supersymmetric standard model requires the presence of two Higgs doublets to
cancel gauge anomalies and to generate masses for both up- and down-type
quarks \cite{Carena:2000yx}.  There are many other models that employ an
extended Higgs sector as well.  The coupling of the Higgs boson to other
particles will be measured by studying the various production processes and
decay modes of the Higgs boson \cite{ATLAS,Zeppenfeld:2000td}.

\begin{figure}[tb]
\begin{center}
\vspace*{0cm} \hspace*{0cm} \epsfxsize=5cm \epsfbox{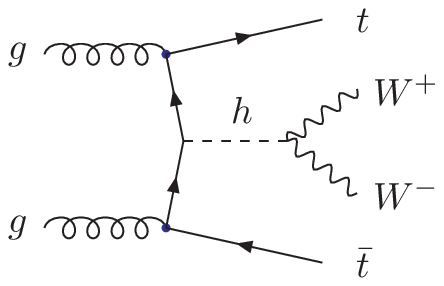} \vspace*{0cm}
\end{center}
\caption{A representative Feynman diagram for associated production of the
Higgs boson and a $t\bar t$ pair, followed by the decay $h\to W^+W^-$.}
\label{fig:ggtth}
\end{figure}

In this paper we study the feasibility of measuring the Yukawa coupling of the
Higgs boson to the top quark via the associated production of the Higgs boson
with a top-quark pair ($t\bar th$) \cite{Ng:1984jm,Kunszt:1984ri}, followed by
the decay $h\to W^+W^-$
\cite{Lee:1977yc,Rizzo:1980gz,Keung:1984hn},\footnote{For $m_h \le 2M_W$, one
or both of the $W$ bosons is virtual.} as shown in Fig.~\ref{fig:ggtth}.  This
decay mode has a branching ratio in excess of $10\%$ for $m_h \gtap 120$ GeV,
and is the dominant decay mode of the Higgs boson for $m_h \gtap 135$ GeV
\cite{Carena:2000yx}.  We study this process at the LHC ($\sqrt s =14$ TeV)
with both low-luminosity and high-luminosity running.  We also consider a
$\sqrt s=$ 100 TeV Very Large Hadron Collider (VLHC) in order to judge the
merits of increased energy. We omit a study at the Tevatron since, even with 30
fb$^{-1}$ of integrated luminosity, the number of signal events is less than
unity once branching ratios are included.

Let us compare the measurement of the top-quark Yukawa coupling via $t\bar th$,
$h\to W^+W^-$ with other methods for measuring this Yukawa coupling.  A less
direct way to measure the top-quark Yukawa coupling at a hadron collider is to
produce the Higgs boson via gluon fusion \cite{Georgi:1978gs}. In the standard
model this process is dominated by a top-quark loop, but if there are other
heavy colored particles that couple to the Higgs boson (such as squarks), they
too contribute to the amplitude, complicating the extraction of the top-quark
Yukawa coupling.  The process $gg\to h$, $h\to W^+W^-$, will be accessible at
the LHC \cite{ATLAS,Glover:1988fn,Barger:1990mn,Dittmar:1996ss}, and perhaps
also at the Tevatron \cite{Han:1998ma,Han:1998sp}, for some range of
Higgs-boson masses. The ratio of $gg\to h$, $h\to W^+W^-$ and $t\bar th$, $h\to
W^+W^-$ is a good probe of additional contributions to the $gg\to h$ amplitude
beyond that of the top quark, as $BR(h\to W^+W^-)$ and many systematic
uncertainties cancel.

An $e^+e^-$ linear collider of sufficient energy and luminosity could also
measure the top-quark Yukawa coupling via $t\bar th$, $h\to W^+W^-$.  A machine
of energy somewhat greater than $m_h + 2m_t$ would be required in order to
overcome the phase space suppression near threshold.  In particular, a $\sqrt
s = 500$ GeV machine would not be adequate for Higgs-boson masses above 130
GeV.  We are not aware of any studies of this process at a linear collider.

For a Higgs boson of mass $\ltap 135$ GeV, the dominant decay mode is $h\to
b\bar b$. Such is the case for the lightest Higgs boson of the minimal
supersymmetric standard model, which has a mass less than about 130 GeV
\cite{Carena:2000yx,Haber:1991aw,Okada:1991vk,Ellis:1991nz}.  The measurement
of the top-quark Yukawa coupling via $t\bar th$ in this decay mode at the LHC
has been studied elsewhere
\cite{ATLAS,Dai:1993gm,Beneke:2000hk,Drollinger:2001ym}.\footnote{This process
might also be accessible at the Tevatron \cite{Goldstein:2001bp}, but only a
crude measurement of the Yukawa coupling could be made.}  The ratio of $t\bar
th$, $h\to b\bar b$ and $t\bar th$, $h\to W^+W^-$ is a good measure of the
ratio of the Yukawa coupling of the $b$ quark and the coupling of the Higgs
boson to $W$ bosons, as $\sigma(t\bar th)$ and many systematic uncertainties
cancel.

In Section~\ref{sec:signal} we discuss the signal for $t\bar th$, $h\to
W^+W^-$, as well as the backgrounds.  We consider final states with two, three,
and four leptons.  In Section~\ref{sec:calc} we discuss the details of the
calculations.  In Section~\ref{sec:results} we present the numerical results
for the number of signal and background events and discuss uncertainties. We
draw conclusions in Section~\ref{sec:conclusions}.

\section{Signal and backgrounds}
\label{sec:signal}

We assume that the Higgs boson has been discovered and its branching ratio
into $W^+W^-$ is known (or is assumed to have the standard-model value).  The
extraction of the top-quark Yukawa coupling then requires a measurement of the
cross section for $t\bar th$, $h\to W^+W^-$.  The total cross section is given
in Table~\ref{tab:totalcs} for a variety of Higgs-boson masses.

\begin{table}[p]
\caption{Total cross section (fb) for $t\bar th$, $h\to W^+W^-$ at the LHC and
a VLHC of $\sqrt{s}=100$ TeV.}
\medskip
\addtolength{\arraycolsep}{0.1cm}
\renewcommand{\arraystretch}{1.4}
\begin{center} \begin{tabular}[4]{c|ccc}
\hline \hline
& \multicolumn{3}{c}{$\sigma(t\bar th)BR(h\to W^+W^-)$} \\[1pt]
$m_h$ (GeV) & 130 & 160 & 190
\\
\hline
LHC   &  110   &   180 & 90  \\[7pt]
VLHC  & 6800   & 12200 & 6700   \\[7pt]
\hline \hline
\end{tabular}
\end{center}
\label{tab:totalcs}
\end{table}

After the top quarks decay, the final state is $W^+W^-W^+W^-b\bar b$.  This
can be divided into different cases, depending on how many $W$ bosons decay
leptonically.  To reduce backgrounds that do not contain top quarks, we
require that both $b$ jets are tagged, with an efficiency $\epsilon_b =$ 60\%
(50\% at high luminosity).\footnote{Since we require two $b$ tags, any
improvement in the $b$-tagging efficiency would be magnified.}  We also include
a lepton identification and reconstruction efficiency $\epsilon_\ell =$ 85\%.
We do not require that either of the top quarks are reconstructed, since the
principal backgrounds also contain top quarks. For a similar reason, we do not
require any missing transverse momentum.  We also do not require that the
Higgs boson be reconstructed, which would be difficult due to the loss of
neutrinos from leptonic $W$ decays. A trigger lepton of $p^T>$ 20 GeV (30 GeV
at high luminosity) is required. The cuts made to simulate the acceptance and
resolution of the detector are given in Table~\ref{tab:detectorcuts}.

\begin{table}[p]
\caption{Cuts applied to simulate the acceptance and resolution of the
detector at the LHC at low luminosity and at high luminosity (in
parentheses).  The high-luminosity cuts are also used for the VLHC.}
\medskip
\addtolength{\arraycolsep}{0.1cm}
\renewcommand{\arraystretch}{1.4}
\begin{center}
\begin{tabular}[4]{ccc}
\hline\hline
$j$    &$p^T>   15 \,(30)$ GeV& $|\eta|<   4.5$\\[5pt]
$b$    &$p^T>   15 \,(30)$ GeV& $|\eta|<   2.5$\\[5pt]
$l$    &$p^T>   10$ GeV       & $|\eta|<   2.5$\\
\hline
 \multicolumn{3}{c}{Trigger lepton: $p^T >  20 \,(30)$ GeV}\\[5pt]
 \multicolumn{3}{c}{$\Delta R_{ij}>0.4$}\\[5pt]
\hline
\hline
\end{tabular}
\end{center}
\label{tab:detectorcuts}
\end{table}

We consider three different cases for the final state, depending on whether
there are two, three, or four leptonic decays of the $W$ bosons in the signal
$W^+W^-W^+W^-b\bar b$.  Each case is discussed separately below.

\subsection{$\ell^\pm\ell^\pm$}

First consider the case where two $W$ bosons decay leptonically.  The signal
consists of $2\ell + 2b + 4j$.  To eliminate the background
$t\bar t + 4j$, we require that the leptons have the same sign. The
backgrounds are:
\begin{itemize}
\item $t\bar tW^\pm jj$, where the $W$ decays leptonically and the top quark
of the same sign decays semileptonically.  We require that the di-jet invariant
mass lie within 25 GeV of the $W$ mass.  This captures most of the signal
events.
\item $t\bar t\ell^+\ell^-jj$, where one lepton is missed, and the top quark
of the same sign as the detected lepton decays semileptonically.  A lepton is
considered missed if it has $p^T<10$ GeV or $|\eta|>2.5$.\footnote{If the
missed lepton has 1 GeV $<p^T<10$ GeV and is within a cone of $\Delta R<0.2$
from a detected lepton, then the detected lepton does not pass the isolation
criteria and the event is rejected.}
\item $t\bar tW^+W^-$, where one $W$ decays leptonically and a top quark of the
same sign decays semileptonically.
\item $t\bar tt\bar t$, where the leptons come from the semileptonic decay of
same-sign top quarks.   We demand that two and only two jets are $b$ tagged.
We accept events with four or more non-$b$-tagged jets.
\end{itemize}

\subsection{$3\ell$}

Next consider the case where three $W$ bosons decay leptonically. The signal
is $3\ell + 2b + 2j $.  Similar backgrounds are present as in
the two-lepton case:
\begin{itemize}
\item $t\bar tW^\pm jj$, where the $W$ and both top quarks decay
semileptonically. We require that the di-jet invariant mass lie within 25 GeV
of the $W$ mass.
\item $t\bar t\ell^+\ell^-$, where both leptons are detected and one of the
top quarks decays semileptonically.  We suppress this background by requiring
that there are no same-flavor, opposite-sign dileptons within 10 GeV of the
$Z$ mass.
\item $t\bar tW^+W^-$, where both $W$'s decay leptonically and one top quark
decays semileptonically, or both top quarks decay semileptonically and one $W$
decays leptonically.
\item $t\bar tt\bar t$, where three of the top quarks decay semileptonically.
We demand that two and only two jets are $b$ tagged. We accept events with two
or more non-$b$-tagged jets.
\end{itemize}

\subsection{$4\ell$}

Finally, consider the case where all four $W$ bosons decay leptonically.  The
signal is $4\ell + 2b $.  The same backgrounds are present
as in the three-lepton case, with the exception of $t\bar tW^\pm jj$:
\begin{itemize}
\item $t\bar t\ell^+\ell^-$, where both leptons are detected and both
top quarks decays semileptonically.  We suppress this background by requiring
that there are no same-flavor, opposite-sign dileptons within 10 GeV of the
$Z$ mass.
\item $t\bar tW^+W^-$, where both $W$'s decay leptonically and both top quarks
decay semileptonically.
\item $t\bar tt\bar t$, where all four top quarks decay semileptonically.
We demand that two and only two jets are $b$ tagged. We accept events with
zero or more non-$b$-tagged jets.
\end{itemize}

There will also be backgrounds from processes not involving top quarks, such
as non-resonant $W^\pm W^\pm+6j$ and $W^\pm Z+6j$ (with one of the leptons from
the $Z$ decay missed) in the $\ell^\pm \ell^\pm$ final state, and $W^\pm
\ell^+\ell^-+4j$ in the $3\ell$ final state. We expect such backgrounds to be
small due to the requirement of double $b$ tagging, in the same way that
non-resonant $W^\pm +4j$ is a small background to $t\bar t$ (with double $b$
tagging) \cite{Beneke:2000hk}. Any such background could be suppressed by
demanding the reconstruction of a top quark in the final state.

Requiring one or more $b$ tags instead of two yields a significant increase in
the number of signal events.  However, non-resonant backgrounds such as those
described above may become significant.  We refrain from pursuing such an
analysis in this paper.

\section{Calculations}
\label{sec:calc}

All calculations are performed at leading order with the code MADGRAPH
\cite{Stelzer:1994ta}, using the CTEQ4L parton distribution functions
\cite{Lai:1997mg} and $\alpha_s(M_Z) = 0.119$ with one-loop running.\footnote{
The factorization and renormalization scales are chosen as follows: i) $t\bar
th$: $\mu_F = \mu_R =(2m_t+m_h)/2$; ii) $t\bar tW^\pm jj$: $\mu_F =
(2m_t+M_W)/2$, $\mu_R = \mu_F$ for two factors of $\alpha_s$ and $p^T$ of each
radiated jet for each of the other two factors; iii) $t\bar t\ell^+\ell^-$:
$\mu_F = \mu_R = (2m_t+m_{\ell\ell})/2$; iv) $t\bar tW^+W^-$: $\mu_F = \mu_R =
m_t+M_W$; v) $t\bar tt\bar t$: $\mu_F = \mu_R = 2m_t$.  For $t\bar th$, the
one-loop running top-quark Yukawa coupling is evaluated at $\mu_R = m_h$.} The
code HDECAY (Version 2.0) \cite{Djouadi:1998yw} is used to calculate $BR(h\to
W^+W^-)$, except for $m_h>2 M_W$, where HDECAY uses the narrow-width
approximation for the $W$ bosons.  We find that $\Gamma(h\to l\bar\nu\bar
l\nu)$ is less than the narrow-width approximation $\Gamma(h\to
W^+W^-)BR(W^-\to l\bar\nu)BR(W^+\to\bar l\nu)$, by as much as 9\% near
threshold.  For $m_h=130,160,190$ GeV, $BR(h\to W^+W^-)=30,92,79\%$,
respectively.

Spin correlations between production and decay processes are maintained in all
calculations.  The top quarks are constrained to be nearly on-shell, but the
$W$ bosons can be off shell.  We also checked each calculation by using the
narrow-width approximation for the top quarks and ignoring the spin correlation
between production and decay.  Good agreement was obtained.

The background $t\bar tW^\pm jj$ is sufficiently complicated that it requires
approximation.  We first calculate the ratio $t\bar tW^\pm jj/t\bar tW^\pm$
for stable top quarks and $W$ bosons, with the acceptance cuts of
Table~\ref{tab:detectorcuts} applied to the two jets.  We then calculate
$t\bar tW^\pm$ with all particles decaying to final-state quarks and leptons,
and multiply by this ratio.

There are subtleties in the calculation of the background $t\bar
t\ell^+\ell^-jj$ where one lepton is missed, and the top quark of the same
sign as the detected lepton decays semileptonically.  First, one must take
care to implement the top-quark width in a manner than ensures gauge
invariance of the amplitude.  We use the ``overall factor scheme''
\cite{Kauer:2001sp}.  Second, the lepton mass must be maintained in the
calculation to regulate the divergence when the missed lepton is collinear
with an observed lepton.\footnote{This is only relevant if the missed lepton
has $p_T < 1$ GeV; otherwise, the observed lepton does not pass the isolation
cut mentioned in a previous footnote.}  We used the muon mass to perform the
calculation; to obtain the result for electrons, we scale the cross section by
a factor $\ln(m_\mu/m_e)$ since the divergence is logarithmic.  This avoids
problems with gauge invariance at very low dilepton invariant mass.  Third, we
approximate $t\bar t\ell^+\ell^-jj$ by calculating $t\bar t\ell^+\ell^-$ with
all particles decaying to final-state quarks and leptons, and then multiplying
by the ratio $t\bar t\ell^+\ell^- jj/t\bar t\ell^+\ell^-$.  Rather than
calculating this ratio, we approximate it from the ratio $t\bar tW^\pm jj/t\bar
tW^\pm$ calculated above, since these two ratios entail QCD radiation from
very similar subprocesses.

In the $t\bar tt\bar t$ background we demand two and only two $b$-tagged jets.
This helps reduce this background, which has four $b$ quarks in the final
state.  If only two of these $b$ quarks are within the acceptance of the
detector, we multiply by $\epsilon_b^2$; if three are within the acceptance,
we multiply by $3\epsilon_b^2(1-\epsilon_b)$; and if all four are within the
acceptance, we multiply by $6\epsilon_b^2(1-\epsilon_b)^2$.

\section{Results}
\label{sec:results}

The number of signal and background events with 30 fb$^{-1}$ of integrated
luminosity at the LHC are given in Table~\ref{tab:backgroundslow} for $m_h =
130, 160, 190$ GeV.  The low-luminosity cuts of Table~\ref{tab:detectorcuts}
have been applied. The two-, three-, and four-lepton final states are
considered individually.  In each case, the signal-to-background ratio is of
order unity, but the number of signal events is not large.  The dominant
backgrounds are $t\bar tW^\pm jj$ and $t\bar t\ell^+\ell^-$; the background
$t\bar tW^+W^-$ is comparatively small, and $t\bar tt\bar t$ is negligible.

We give in Table~\ref{tab:backgroundshigh} the number of signal and background
events with 300 fb$^{-1}$ of integrated luminosity at the LHC.  The
high-luminosity cuts of Table~\ref{tab:detectorcuts} have been applied.  The
number of events is increased due to the factor of ten increase in integrated
luminosity, but this is partially compensated by the decreased acceptance.

Table~\ref{tab:backgroundsvlhc} gives the number of events at a 100 TeV VLHC
with 300 fb$^{-1}$ of integrated luminosity.\footnote{Preliminary results for
the VLHC were reported in Ref.~\cite{Baur:2002ka}.} The high-luminosity cuts of
Table~\ref{tab:detectorcuts} have been applied. The increased energy results
in many more events than at the LHC with the same integrated luminosity. The
signal-to-background ratio remains of order unity.   All four sources of
background are comparable at the VLHC.

\begin{table}[p]
\caption{The number of signal and background events in 30 fb$^{-1}$ of
integrated luminosity at the LHC for $t\bar th$, $h\to W^+W^-$ in the two-,
three-, and four-lepton final states. The low-luminosity cuts of
Table~\ref{tab:detectorcuts} are applied. The background $t\bar t\ell^+\ell^-$
has two additional jets in the dilepton case only. Throughout the table, $\ell$
denotes a summation over $e$ and $\mu$. $B$ is the total number of background
events.}

\addtolength{\arraycolsep}{0.1cm}
\renewcommand{\arraystretch}{1.2}
\medskip
\begin{center} \begin{tabular}[4]{c|ccc|cccc|c}
\hline \hline & \multicolumn{3}{c|}{$t\bar th$} &
\multicolumn{4}{c|}{backgrounds}
\\[1pt]
$m_h$ (GeV) & 130 & 160 & 190 & $t\bar tW^\pm jj$ & $t\bar t\ell^+\ell^-(jj)$
& $t\bar tW^+W^-$ & $t\bar tt\bar t$ & $B$
\\
\hline
$2\ell$ & 6.4 & 15 & 8.3& 10 & 1.9& 0.86 & 1.6 & 14\\[7pt]
$3\ell$ & 3.8 & 8.8 &  4.7& 2.4& 4.9& 0.49&0.75& 8.5\\[7pt]
$4\ell$ & 0.38& 0.67& 0.34& --- & 0.67 & 0.036& 0.009 & 0.72 \\[7pt]
\hline \hline
\end{tabular}
\end{center}
\label{tab:backgroundslow}
\end{table}
\begin{table}[p]
\caption{Same as Table~\ref{tab:backgroundslow}, except the high-luminosity
cuts of Table~\ref{tab:detectorcuts} are applied, and 300 fb$^{-1}$ of
integrated luminosity are collected at the LHC.}

\addtolength{\arraycolsep}{0.1cm}
\renewcommand{\arraystretch}{1.2}
\medskip
\begin{center} \begin{tabular}[4]{c|ccc|cccc|c}
\hline \hline & \multicolumn{3}{c|}{$t\bar th$} &
\multicolumn{4}{c|}{backgrounds} &
\\[1pt]
$m_h$ (GeV)& 130 & 160 & 190 & $t\bar tW^\pm jj$ & $t\bar t\ell^+\ell^-(jj)$ &
$t\bar tW^+W^-$ & $t\bar tt\bar t$ & $B$
\\
\hline
$2\ell$ & 8.1 & 24  & 16  & 19  & 3.2  &2.1 & 4.2 & 29 \\[7pt]
$3\ell$ & 12  & 27  & 16  &  4.6& 17   &1.8 & 3.6 & 27 \\[7pt]
$4\ell$ &2.1 & 3.8 & 2.0 & --- & 3.9   &0.21& 0.20& 4.3\\[7pt]
\hline \hline
\end{tabular}
\end{center}
\label{tab:backgroundshigh}
\end{table}
\begin{table}[p]
\caption{Same as Table~\ref{tab:backgroundshigh}, except at the VLHC ($\sqrt s
= 100$ TeV). The high-luminosity cuts of Table~\ref{tab:detectorcuts} are
applied, and 300 fb$^{-1}$ of integrated luminosity are collected.}

\addtolength{\arraycolsep}{0.1cm}
\renewcommand{\arraystretch}{1.2}
\medskip
\begin{center} \begin{tabular}[4]{c|ccc|cccc|c}
\hline \hline & \multicolumn{3}{c|}{$t\bar th$} &
\multicolumn{4}{c|}{backgrounds} &
\\[1pt]
$m_h$ (GeV) & 130 & 160 & 190 & $t\bar tW^\pm jj$ & $t\bar t\ell^+\ell^-(jj)$
& $t\bar tW^+W^-$ & $t\bar tt\bar t$ & $B$
\\
\hline
$2\ell$ & 370& 1100& 840& 440& 360& 140& 1065 & 2005\\[7pt]
$3\ell$ & 500& 1200& 780& 100& 640& 130& 640  & 1510\\[7pt]
$4\ell$ & 85 & 160 &  91& --- &140 & 15&  72  & 227 \\[7pt]
\hline \hline
\end{tabular}
\end{center}
\label{tab:backgroundsvlhc}
\end{table}

\begin{figure}[p]
\begin{center}
\vspace*{-2cm}
\hspace*{-.5cm}
\epsfxsize=13cm     
\epsfbox{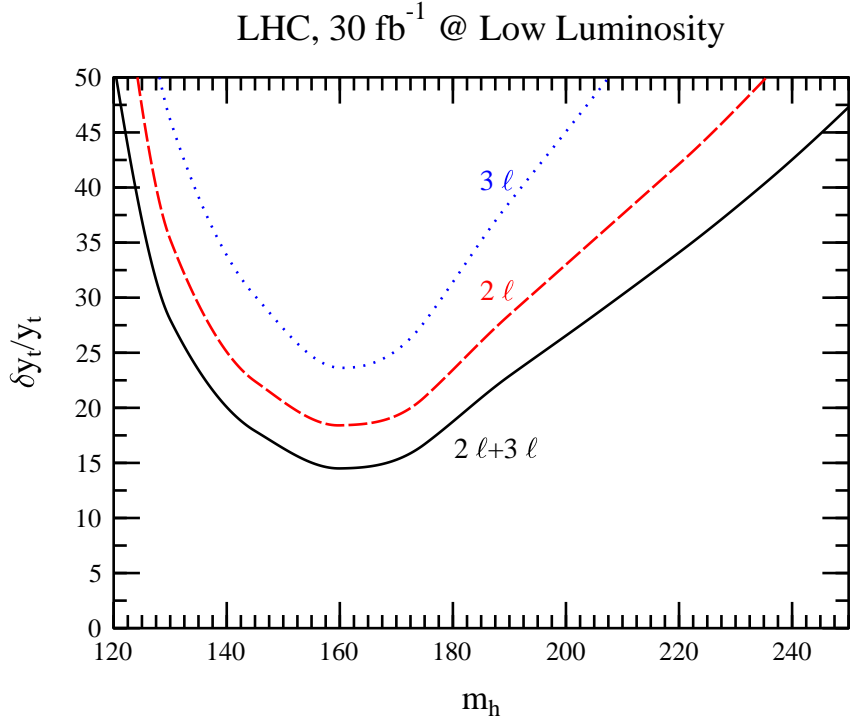}
\vspace*{-.5cm}
\end{center}
\caption{The statistical uncertainty in the top-quark Yukawa coupling, $\delta
y_t/y_t$, from $t\bar th$, $h\to W^+W^-$ at the low-luminosity LHC (30
fb$^{-1}$ of integrated luminosity).  Results for the two- and three-lepton
final states, as well as their sum, are shown.} \label{fig:precisionlow}
\end{figure}

\begin{figure}[p]
\begin{center}
\vspace*{-1cm}
\hspace*{-.5cm}
\epsfxsize=13cm     
\epsfbox{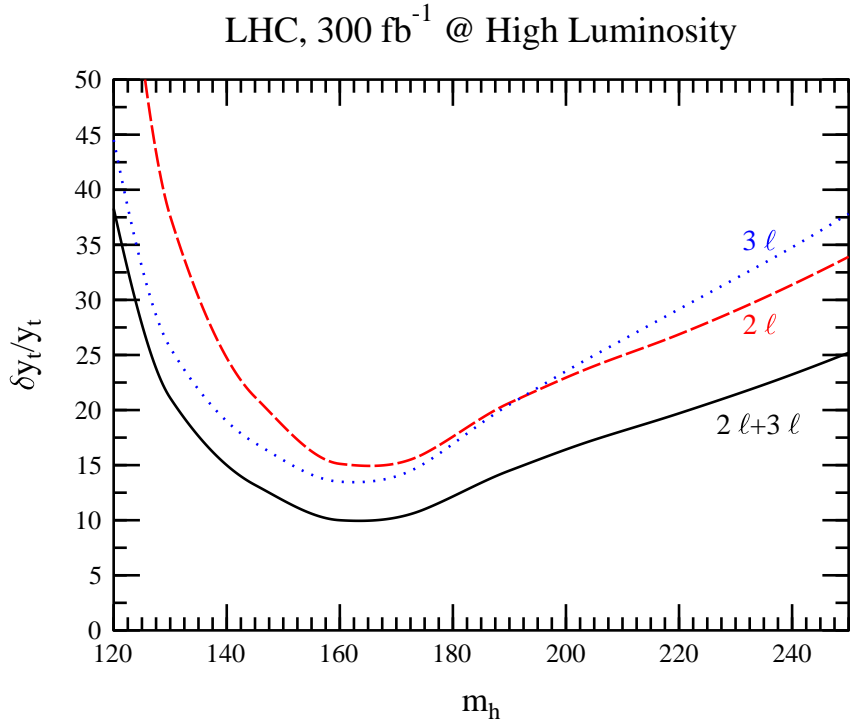}
\vspace*{-.5cm}
\end{center}
\caption{Same as Fig.~\ref{fig:precisionlow}, but at the high-luminosity LHC
(300 fb$^{-1}$ of integrated luminosity).} \label{fig:precisionhigh}
\end{figure}

There are three sources of uncertainty in the extraction of the top-quark
Yukawa coupling from the measured cross section: statistical, systematic, and
theoretical.  The statistical uncertainty in the measured cross section is
$\sqrt{S+B}/S$, where $S$ and $B$ are the number of signal and background
events, respectively.  Since the cross section is proportional to the square
of the Yukawa coupling, the statistical uncertainty in the measurement of the
Yukawa coupling is half that of the cross section.  We show in
Figs.~\ref{fig:precisionlow} and \ref{fig:precisionhigh} the statistical
uncertainty $\delta y_t/y_t$ at the low- and high-luminosity LHC, respectively.
The statistical uncertainty from the two- and three-lepton final states are
given separately, as well as the statistical uncertainty from the sum of the
two channels, combined in quadrature. The statistical uncertainty in the
four-lepton final state is much greater, due to the dearth of signal events.
Combining the low- and high-luminosity runs at the LHC in quadrature yields a
statistical uncertainty in the measurement of the Yukawa coupling of $16\%,
8\%, 12\%$ for $m_h = 130, 160, 190$ GeV, respectively.

The systematic uncertainty arises from our ability to measure and calculate
the backgrounds.  Since the signal is not separated from the backgrounds by
kinematics, an accurate knowledge of the backgrounds is essential.  It is
beyond the scope of this work to attempt to estimate the systematic
uncertainty in the measurement of the backgrounds; an uncertainty of $20\%$ or
less is desired in order to match the statistical uncertainty.

The dominant background in the three- and four-lepton final states is $t\bar
t\ell^+\ell^-$.   This background could be estimated by measuring $t\bar tZ$,
$Z\to \ell^+\ell^-$, and extrapolating away from the $Z$ resonance.  This
background is absent altogether if all opposite-sign leptons are of different
flavor (e.g., $\mu^+\mu^+e^-$ or $\mu^+\mu^+e^-e^-$), but the number of events
of this type is sufficiently small that the statistical uncertainty in the
cross section increases when one considers only these events.

The dominant background in the two-lepton final state is $t\bar tW^\pm jj$. It
is likely that our calculation overestimates this background, because, using
the cuts of Table~\ref{tab:detectorcuts}, each of the two jets can lie near the
soft and collinear regions of phase space where perturbation theory overshoots
the true cross section \cite{Rainwater:1996vq}.  This background could be
estimated by measuring the cross section for $W^\pm jj$ invariant masses far
from the Higgs mass, and then extrapolating.  This is nontrivial, as it
assumes that one can correctly identify the $W^\pm jj$ system with good
efficiency.

The theoretical uncertainty stems from the uncertainty in the subprocess cross
section $gg,q\bar q\to t\bar th$ and the uncertainty in the parton
distribution functions.  The subprocess cross section has been calculated at
next-to-leading order in the strong interaction \cite{Beenakker:2001rj}. The
dependence of the cross section on the common factorization and
renormalization scales is relatively mild at the LHC, varying by about $10\%$
as the scale is varied from half to twice its central value of
$\mu_R=\mu_F=(2m_t+m_h)/2$.  The largest uncertainty in the parton
distribution functions is from the gluons. The gluon-gluon luminosity is known
to about $10\%$ accuracy at the LHC \cite{Huston:1998jj}, and will be measured
via $gg\to t\bar t$. Hence the total theoretical uncertainty in the cross
section is about $15\%$.  This is adequate in comparison with the statistical
uncertainty.

The statistical uncertainty in $\delta y_t/y_t$ at the VLHC from the sum of
the two- and three-lepton final states is $3.3\%, 1.6\%, 2.2\%$ for $m_h = 130,
160, 190$ GeV. This is negligible compared with the likely systematic
uncertainty. Thus the measurement of the top-quark Yukawa coupling at the VLHC
would be limited almost entirely by the systematic uncertainty.

The background $t\bar tt\bar t$ is negligible at the LHC, but it is
significant at the VLHC; it is the largest background in the two-lepton final
state, and one of the two largest backgrounds in the three-lepton final state.
This background has two additional jets in the final state compared with the
signal, so it could be reduced by vetoing events with extra jets.  For example,
vetoing events with two extra jets (but not one) reduces the $t\bar tt\bar t$
background at the VLHC to 500, 170, 21 in the two-, three, and four-lepton
final states, respectively.  While this does not decrease the statistical
uncertainty (which is negligible in any case), it may be relevant to control
the systematic uncertainty.

The process $t\bar th$, $h\to ZZ$ may also be accessible at the VLHC.  If both
$Z$ bosons decay leptonically, the Higgs-boson mass could be reconstructed,
which would reduce the backgrounds.


\section{Conclusions}
\label{sec:conclusions}

We have studied the measurement of the top-quark Yukawa coupling to the Higgs
boson via $t\bar th$, $h\to W^+W^-$ at the LHC.  A signal is available in the
two- and three-lepton final states, and the signal to background ratio is of
order unity.  Combining both channels, and assuming 30 fb$^{-1}$ of integrated
luminosity at low luminosity and 300 fb$^{-1}$ at high luminosity, we find a
statistical uncertainty in the measurement of the Yukawa coupling $\delta
y_t/y_t$ of $16\%, 8\%, 12\%$ for $m_h = 130, 160, 190$ GeV, respectively.

The statistical uncertainty in the Yukawa coupling would be negligible at a
100 TeV VLHC.  Thus the uncertainty in the Yukawa coupling would be dominated
by the systematic uncertainty.

\section*{Acknowledgments}

\indent\indent We are grateful to T.~Stelzer for extensive assistance and
advice.  We are also grateful for conversations and correspondence with
U.~Baur, M.~Dittmar, K.~Jacobs, T.~Liss, S.~Nikitenko, K.~Pitts, J.~Strologas,
and G.-P.~Yeh.  We thank the organizers of the Snowmass 2001 workshop and the
Aspen Center for Physics, where part of this work was conducted.  This work was
supported in part by the U.~S.~Department of Energy under contract
No.~DOE~DE-FG02-91ER40677.


\begin{thebibliography}{99}

\bibitem{:2001xw} LEP Higgs Working Group for Higgs boson searches
Collaboration, LHWG-NOTE-2001-03,
arXiv:hep-ex/0107029.

\bibitem{Abbaneo:2001ix}
D.~Abbaneo {\it et al.}  [ALEPH Collaboration],
arXiv:hep-ex/0112021.

\bibitem{Carena:2000yx}
M.~Carena {\it et al.}, ``Report of the Tevatron Higgs working group,''
hep-ph/0010338.

\bibitem{ATLAS} ATLAS Technical Design Report, Vol.~II, CERN/LHCC/99-15 (1999).

\bibitem{CMS} CMS Collaboration, Technical Proposal, CERN/LHCC/94-38 (1994).

\bibitem{Zeppenfeld:2000td}
D.~Zeppenfeld, R.~Kinnunen, A.~Nikitenko and E.~Richter-Was,
Phys.\ Rev.\ D {\bf 62}, 013009 (2000) [arXiv:hep-ph/0002036].

\bibitem{Ng:1984jm}
J.~N.~Ng and P.~Zakarauskas,
Phys.\ Rev.\ D {\bf 29}, 876 (1984).

\bibitem{Kunszt:1984ri}
Z.~Kunszt,
Nucl.\ Phys.\ B {\bf 247}, 339 (1984).

\bibitem{Lee:1977yc}
B.~W.~Lee, C.~Quigg and H.~B.~Thacker,
Phys.\ Rev.\ Lett.\  {\bf 38}, 883 (1977).

\bibitem{Rizzo:1980gz}
T.~G.~Rizzo,
Phys.\ Rev.\ D {\bf 22}, 722 (1980).

\bibitem{Keung:1984hn}
W.~Y.~Keung and W.~J.~Marciano,
Phys.\ Rev.\ D {\bf 30}, 248 (1984).

\bibitem{Georgi:1978gs}
H.~M.~Georgi, S.~L.~Glashow, M.~E.~Machacek and D.~V.~Nanopoulos,
Phys.\ Rev.\ Lett.\ {\bf 40}, 692 (1978).

\bibitem{Glover:1988fn}
E.~W.~Glover, J.~Ohnemus and S.~S.~Willenbrock,
Phys.\ Rev.\ D {\bf 37}, 3193 (1988).

\bibitem{Barger:1990mn}
V.~D.~Barger, G.~Bhattacharya, T.~Han and B.~A.~Kniehl,
Phys.\ Rev.\ D {\bf 43}, 779 (1991).

\bibitem{Dittmar:1996ss}
M.~Dittmar and H.~Dreiner,
Phys.\ Rev.\ D {\bf 55}, 167 (1997) [arXiv:hep-ph/9608317].

\bibitem{Han:1998ma}
T.~Han and R.~J.~Zhang,
Phys.\ Rev.\ Lett.\  {\bf 82}, 25 (1999) [arXiv:hep-ph/9807424].

\bibitem{Han:1998sp}
T.~Han, A.~S.~Turcot and R.~J.~Zhang,
Phys.\ Rev.\ D {\bf 59}, 093001 (1999) [arXiv:hep-ph/9812275].

\bibitem{Haber:1991aw}
H.~E.~Haber and R.~Hempfling,
Phys.\ Rev.\ Lett.\  {\bf 66}, 1815 (1991).

\bibitem{Okada:1991vk}
Y.~Okada, M.~Yamaguchi and T.~Yanagida,
Prog.\ Theor.\ Phys.\  {\bf 85}, 1 (1991).

\bibitem{Ellis:1991nz}
J.~Ellis, G.~Ridolfi and F.~Zwirner,
Phys.\ Lett.\ B {\bf 257}, 83 (1991).

\bibitem{Dai:1993gm}
J.~Dai, J.~F.~Gunion and R.~Vega,
Phys.\ Rev.\ Lett.\ {\bf 71}, 2699 (1993) [hep-ph/9306271].

\bibitem{Beneke:2000hk}
M.~Beneke {\it et al.}, ``Top quark physics,'' hep-ph/0003033, in
``Proceedings of the Workshop on Standard Model Physics at the LHC,''
eds.~G.~Altarelli and M.~Mangano, CERN 2000-004.

\bibitem{Drollinger:2001ym}
V.~Drollinger, T.~Muller and D.~Denegri,
arXiv:hep-ph/0111312.

\bibitem{Goldstein:2001bp}
J.~Goldstein, C.~S.~Hill, J.~Incandela, S.~Parke, D.~Rainwater and D.~Stuart,
Phys.\ Rev.\ Lett.\  {\bf 86}, 1694 (2001) [hep-ph/0006311].

\bibitem{Stelzer:1994ta}
T.~Stelzer and W.~F.~Long,
Comput.\ Phys.\ Commun.\ {\bf 81}, 357 (1994)
[hep-ph/9401258].

\bibitem{Lai:1997mg}
H.~L.~Lai {\it et al.},
Phys.\ Rev.\ D {\bf 55}, 1280 (1997) [hep-ph/9606399].

\bibitem{Djouadi:1998yw}
A.~Djouadi, J.~Kalinowski and M.~Spira,
Comput.\ Phys.\ Commun.\ {\bf 108}, 56 (1998)
[hep-ph/9704448].

\bibitem{Kauer:2001sp}
N.~Kauer and D.~Zeppenfeld,
Phys.\ Rev.\ D {\bf 65}, 014021 (2002) [arXiv:hep-ph/0107181].

\bibitem{Baur:2002ka}
U.~Baur {\it et al.},
arXiv:hep-ph/0201227.

\bibitem{Rainwater:1996vq}
D.~Rainwater, D.~Summers and D.~Zeppenfeld,
Phys.\ Rev.\ D {\bf 55}, 5681 (1997) [arXiv:hep-ph/9612320].

\bibitem{Beenakker:2001rj}
W.~Beenakker, S.~Dittmaier, M.~Kramer, B.~Plumper, M.~Spira and P.~M.~Zerwas,
hep-ph/0107081.

\bibitem{Huston:1998jj}
J.~Huston, S.~Kuhlmann, H.~L.~Lai, F.~I.~Olness, J.~F.~Owens, D.~E.~Soper and
W.~K.~Tung,
Phys.\ Rev.\ D {\bf 58}, 114034 (1998) [arXiv:hep-ph/9801444].

\end{thebibliography}
\end{document}